\documentclass{article}
\usepackage{spconf,amsmath,graphicx}
\usepackage{url}
\usepackage{tabularx} 
\usepackage[numbers,square]{natbib}

\title{An LSTM based architecture to relate speech stimulus to EEG}
\name{\begin{tabular}{c}Mohammad Jalilpour Monesi$^{1,2}$, Bernd Accou$^{1,2}$, Jair Montoya-Martinez$^2$, \\ Tom Francart$^2$, Hugo Van Hamme$^1$\end{tabular}\thanks{ The work is funded by KU Leuven Special Research Fund C24/18/099 (C2 project to Tom Francart and Hugo Van hamme).}\thanks{This project has received funding from the European Research Council (ERC) under the European Union’s Horizon 2020 research and innovation programme (grant agreement No 637424, ERC starting Grant to Tom Francart).}}
\address{$^1$KU Leuven, PSI, Dept. of Electrical engineering (ESAT), Leuven, Belgium\\
$^2$KU Leuven, ExpORL, Dept. Neurosciences, Leuven, Belgium
}

\begin{document}
%
\maketitle
\begin{abstract}
Modeling the relationship between natural speech and a recorded electroencephalogram (EEG) helps us understand how the brain processes speech and has various applications in neuroscience and brain-computer interfaces. In this context, so far mainly linear models have been  used. However, the decoding performance of the linear model is limited due to the complex and highly non-linear nature of the auditory processing in the human brain. We present a novel Long Short-Term Memory (LSTM)-based architecture as a non-linear model for the classification problem of whether a given pair of (EEG, speech envelope) correspond to each other or not. The model maps short segments of the EEG and the envelope to a common embedding space using a CNN in the EEG path and an LSTM in the speech path. The latter also compensates for the brain response delay. In addition, we use transfer learning to fine-tune the model for each subject. The mean classification accuracy of the proposed model reaches 85\%, which is significantly higher than that of a state of the art Convolutional Neural Network (CNN)-based model (73\%) and the linear model (69\%).

\end{abstract}

\section{Introduction}
\label{sec:intro}


Over the past two decades, researchers have tried to model how natural running speech is encoded in the human brain \cite[e.g.][]{ahissar_speech_2001, aiken_human_2008, ding_emergence_2012, crosse_multivariate_2016, vanthornhout_speech_2018}. In an experimental paradigm where natural running speech is presented to a listener while the EEG is recorded, linear regression is used to  either decode features of  the speech signal from the EEG signal (backward model), or to predict the EEG signal from the speech stimulus (forward model). Then, the correlation between the actual and the predicted signal is computed and used as a measure of neural tracking of speech \cite[e.g.][]{ding_emergence_2012, diliberto_low-frequency_2015, crosse_multivariate_2016, vanthornhout_speech_2018}. This method has applications in domains such as audiology, as part of an objective measure of speech intelligibility \cite{vanthornhout_speech_2018, lesenfants_predicting_2019}, as well as other potential applications in neuroscience such as brain computer interfaces (BCIs).

Unfortunately, the correlations between actual and predicted signal with either technique are small (in the order of 0.1), limiting its applicability. This is partly due to the use of simple linear models, which cannot model the complex and dynamic nature of the brain. Another problem of linear models is the delay (lag) between speech and EEG \cite{aiken_human_2008, mirkovic_decoding_2015}. Usually this delay varies across subjects which necessitates the use of subject-specific decoders. But importantly, the delay also varies within-subject during a recording \cite[e.g.][]{ding_adaptive_2013}, depending on their state of mind (attention, arousal, effort, etc), which cannot be modeled with a linear model. Hence, linear models have a high variance in performance, both within and across subjects. Furthermore, linear models need long segments of test data (between 30-60 seconds) \cite{mirkovic_decoding_2015, de_cheveigne_decoding_2018, ding_adaptive_2013, fuglsang_noise-robust_2017}  which is too long for some online applications.

Considering the complex and highly non-linear nature of auditory processing in the human brain, and with
the recent success of deep learning methods in vision related
tasks and in automatic speech recognition, using artificial neural networks (ANNs) in this context seems a worthwhile exploration.  In \cite{de_taillez_machine_2017}, a simple feedforward neural network with hyperbolic tangent activation functions was used to reconstruct the envelope from the EEG. Recently, CNNs have also been applied for auditory attention decoding (AAD), in which case the subject attends to one of two concurrent speakers, and the system decodes the attended speaker \cite{deckers_eeg-based_2018, ciccarelli_comparison_2019}.

In this work, inspired by recent advances in AAD \cite{ciccarelli_comparison_2019, deckers_eeg-based_2018}, we have redefined the more difficult regression problem of reconstructing the speech stimulus from the EEG as a classification problem. The goal is to design a model that can determine whether a given pair of EEG and speech envelope correspond to each other or not. To this end, we have designed a novel LSTM-based deep learning model \cite{hochreiter_long_1997}. When the average performance of the network (\% correct) on this task is high, the performance can be used as a proxy for neural coding of the speech envelope.

\section{Methodology}
\label{sec:pagestyle}

In this section, first, we will explain our data collection and preprocessing step. Then, we will explain the our classification task in detail. Finally, we will present the proposed LSTM model to be used in the classification task. The linear model and the CNN model presented in \cite{ciccarelli_comparison_2019} will be used as baseline and state of the art, respectively, for comparisons.

\subsection{Data collection and preprocessing}

In our protocol, we present natural running speech, which
are stories in quiet, and record the EEG signal simultaneously. 90 normal hearing native Flemish subjects participated in this study. All subjects went through screening for normal hearing test with pure tone
audiometry and the Flemish MATRIX-test \cite{luts2014development}. 

Stories were chosen from a set of ten unique stories of roughly the same length (14 minutes and 30 seconds). However, the number of stories presented to each subject varies between a minimum of 1 and a maximum of 8. The presentation order of  the stories was randomized for each subject. These stories were presented binaurally at 62~dBA with Etymotic ER-3A insert phones. After each stimulus (story), a comprehension question was asked to the subjects to ensure they paid attention. EEG data was recorded using a 64 channel Biosemi Active-Two EEG system at 8~kHz sampling rate. The stimuli were presented using the APEX 4 software platform \cite{francart_apex_2008} developed at ExpORL. The experiments took place in an electromagnetically shielded and soundproofed cabin.

After each recording, the EEG signal is synchronized with the
corresponding stimulus. The envelope of the speech stimulus
is extracted using the ’powerlaw subbands’ method from \cite{biesmans_auditory-inspired_2017}. A
multi channel Wiener filter \cite{somers_generic_2018} is used to remove artefacts from the EEG recordings.
Both EEG and envelope are bandpass filtered between 0.5~Hz and
32~Hz. Then, both EEG and envelope are downsampled to 64
Hz. Finally, mean and variance normalization is applied to the EEG
and envelope for each recording. Furthermore, we divided each subject's recorded data into training, validation and test sets. The training set contains the first and the last 40\% of each recording( 80\% in total) and the remaining 20\% from the middle of the recording is equally divided between the validation and the test sets in this order.

\subsection{Classification task}

We use a 10 seconds time window with 90\% overlap to cut the recorded EEG and envelope into several segments. Each of these segments is considered a sample to the classifiers (models). The positive samples include all the (EEG, envelope) pairs in which the speech envelope corresponds to the recorded EEG. Similarly, the negative samples are pairs of (EEG, envelope), where the speech envelope does not correspond to the recorded EEG. To generate these negative samples (mismatched envelopes), we extract 10 seconds of envelope from the same dataset. This mismatched envelope is chosen randomly either to start one second after the end of matched envelope or to end one second before beginning of the matched envelope. 

Given a pair of (EEG, envelope), we want to have a model (classifier) that can correctly determine whether it is a matched (positive sample) or mismatched (negative sample). Note that compared to a two envelope setup where the task is to choose the matched envelope between the two given candidate envelope (similar to AAD), our one envelope setup is more challenging. We will use classification accuracy as the evaluation measure to compare
performance of the models.

\subsection{Models}

\begin{figure*}[htb!]
	
		\centering
		\centerline{\includegraphics[width=\textwidth, height=4cm]{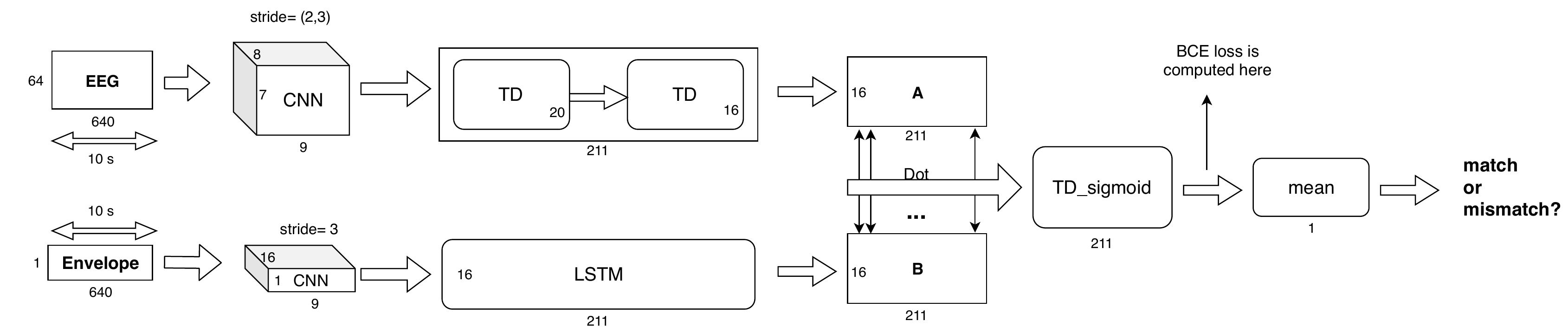}}
	
	%
	\caption{Our proposed LSTM-based model for classification of match/mismatch. TD refers to time distributed which applies a dense layer to every temporal slice of the input. Dot is a layer that applies dot product (cosine similarity) to its two norm one input vectors.}
	\label{fig:res}
\end{figure*}


Our proposed model to classify a pair of (EEG, envelope) as matched or mismatched is shown in Figure \ref{fig:res}. It is composed of two separate networks mapping either the EEG or the speech envelope to a sequence of embedding vectors of unit length. The network training is organised to obtain a common embedding space for EEG and speech envelope. This common space ideally should have a very similar representation (i.e. aligned embedding vectors) when speech envelope and EEG are matched, while dissimilar representation (opposite embedding vectors) are expected for mismatched samples. 

Instead of mapping the whole 10 seconds of EEG (640 time samples) and envelope into the common embedding space, using a CNN layer we divide the whole 10 second segment into shorter sections which are each mapped to the embedding space (columns of matrices A and B in figure~\ref{fig:res}). 
Our hypothesis is that shorter segments of the input data contain enough information, so it should be possible for the network to align EEG and envelope of these shorter segments to each other in the new common space. As a result, in our loss function, we will have more than one sample (i.e. 211) per 10 second segment. We use binary cross entropy (BCE) as our loss function immediately after the cosine similarity scores.

We have opted for an LSTM layer for speech envelope processing (lower part of the network) to also give our network the ability to compensate for the brain response delay and hence synchronize with the recorded EEG. Furthermore, the LSTM can try to mimic the brain's response to speech input. However, an LSTM's memory capability, expressed in recurrence steps, is limited. To address this, we use a CNN preprocessing layer with a stride of 3 to decrease the number of recurrence steps the LSTM has to apply to model the typical brain response delay to less than 10. The EEG signal is processed by a CNN followed by two dense layers. The CNN layer uses the same temporal stride of 3 to maintain the same sampling rate for both of the EEG and the speech envelope.

Our proposed network contains several important hyperparameters including learning rate, number of LSTM units, number of units in the time distributed layers (TD in Figure \ref{fig:res}), size of the CNN kernels, etc. We have used our train and validation sets to find the optimal values for these hyper parameters.
We used 30 epochs with early stopping in the training procedure. The optimal values for the network's hyperparameters are shown in Figure \ref{fig:res}. After hyper parameter tuning, our network has around 8000 trainable parameters. The code for the proposed model is provided in https://github.com/jalilpour-m/match-mismatch\_icassp2020.


We used the linear backward model \cite{osullivan_attentional_2015, crosse_multivariate_2016} as our  linear baseline. To do so, first, we use the linear backward model to reconstruct the envelope from the recorded EEG. Then, we calculate the Spearman correlation between this reconstructed envelope and the candidate envelope. Finally, if the correlation score is above a threshold (tuned for equal false positives and false negatives on the validation set) then it is classified as matched otherwise it is classified as mismatched.

Additionally, we used a CNN-based model proposed in \cite{ciccarelli_comparison_2019} as a state of the art architecture. We will refer to this network as the SoA (state of the art) network. We tuned the hyperparameters of the SoA network for our dataset the same way we did for our LSTM model.

\section{Results}
\label{sec:typestyle}

\subsection{Subject dependence}
The models are trained in three different scenarios. First, we can train one specific model per subject, i.e. the training data of just one subject is used. This is repeated for each subject and results in as many models as there are subjects. We will call this scenario subject dependent (SD). Second, we use all the data available from all the subjects to train one generic model, and test it separately for each individual subject. We will use the term subject independent (SI) to refer to this scenario. In the third case, again we train one specific model for each subject, but this time, we initialize the parameters of each SD model with the SI model. Then, we use the train set of each subject to fine tune the model's layers in the EEG path (upper layers in Figure \ref{fig:res}). The term transfer learning (TL) will be used to refer to this scenario. It is worth noting that in practical applications, we are interested mostly in the SI and TL scenarios, because the TL scenario is a better alternative for the SD scenario. But we may also report the performance of models in the SD scenario for comparison reasons.

\subsection{Effect of data size}
\label{ssec:subhead}

In this section, we analyze the effect of data size on the classification accuracy of the models. Therefore, we applied all three models to three different data sizes. First, we evaluated methods in the SD scenario (least amount of data). Then, we applied the models in the SI scenario with 20 subjects. Finally, models were applied to the SI scenario again, but this time we used all the data available (90 subjects) in the training. Note that we still use the same 20 subjects in the test set. In addition, in order to evaluate how well the models generalize, we used data of 20 different subjects as holdout data.

\begin{figure}[htb]
	
	\begin{minipage}[b]{1\linewidth}
		\centering
		\centerline{\includegraphics[width=\textwidth]{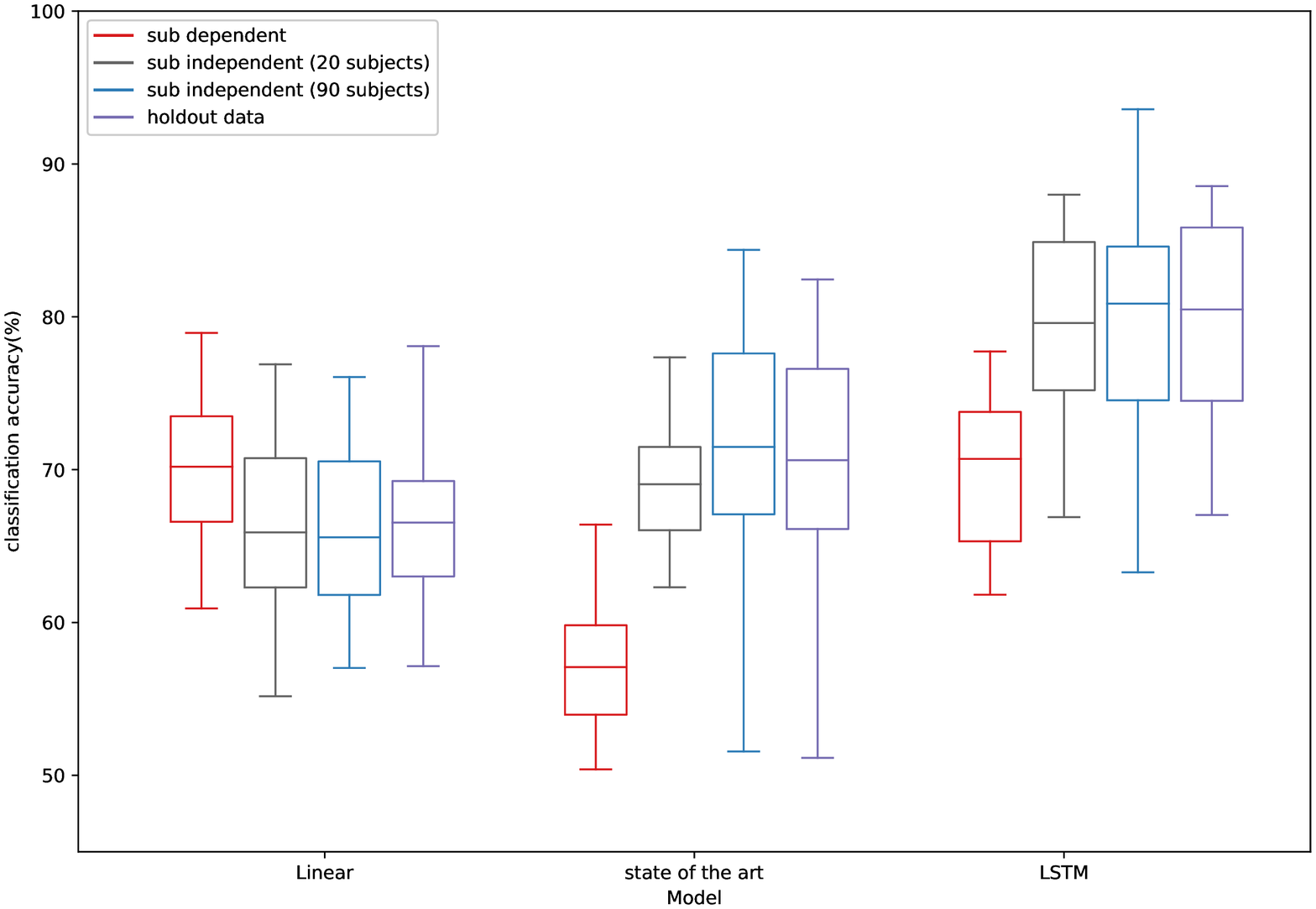}}
	\end{minipage}

	\caption{Effect of data size on classification accuracy: in `sub dependent', models are trained on one subject only. In `sub independent', one generic model is trained using data of all subjects. Numbers inside parentheses in the legend indicate the number of subjects used in training. Box plots are shown over a fixed test set of 20 subjects for the first three scenarios. In 'holdout data', the generic model is evaluated on data of 20 new subjects.}
	\label{fig:big_data}
\end{figure}

Box plots over the 20 subjects are shown in Figure \ref{fig:big_data}. For the subject dependent case, we see that the simple linear model has the same or even better performance than the two deep-learning-based models. But when the size of the data increases in the SI scenario, the performance of the deep learning models increases while the classification accuracy of linear model decreases. These results are in line with our expectation from deep learning literature that more data helps to obtain an ANN model that generalizes better. In contrast, we see that the linear model cannot benefit from more data with subject variation (20 times more data). The proposed LSTM model's average accuracy reaches 80\% in the SI scenario which outperforms the linear model (66\%) (W = $9$, $p<0.001$) and the SoA model (72\%) (W = $42.5$, $ p=0.01888$). The statistics are reported from a Wilcoxon signed-rank test. Furthermore, we do not see further increase in accuracy when we increase the number of training subjects from 20 to 90. Our results for the holdout data suggest that the models are well generalized over unseen data.

\subsection{Transfer learning}

\begin{figure}[htb]
	
	\begin{minipage}[b]{1\linewidth}
		\centering
		\centerline{\includegraphics[width=\textwidth]{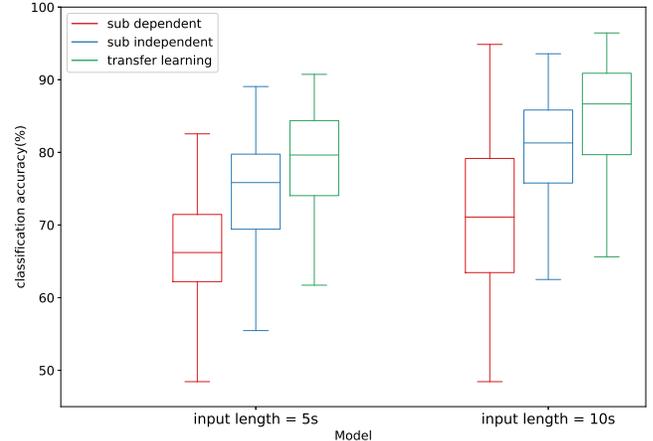}}
	\end{minipage}

	\caption{Classification accuracies of the proposed LSTM model in SD, SI, and TL scenarios for two different lengths of input segments. Box plots are shown over 90 subjects.}
	\label{fig:short_TL}
\end{figure}

Here, we try to investigate the idea of transferring knowledge between ANN models. While more data clearly improves our deep learning models, it is also known that an EEG can be very idiosyncratic, so  individualisation of the models may improve performance. However, it is not possible to collect large amounts of data for one individual subject.  Therefore we train a generic model using all the subjects, then we fine tune it for each subject. Since the speech signal is the same for all subjects, we only retrain layers in the EEG path  (the upper layers in our network). The results are shown in Figure~\ref{fig:short_TL}. For input frames of 10 seconds long, performance significantly improved with TL (85\%) (z =  $-6.97$, $p<0.001$) compared to the classical SD setting (71\%). As performance now approaches 100\%, to avoid ceiling effects, we decreased the input frame length to 5 seconds. Overall, this reduction decreases performance, as expected. In this case we also obtained the best performance in the TL scenario (80\%), which was again significantly better than the SD scenario (66\%) ( z = $-7.50$ , $p<0.001$) and the SI scenario (74\%) ( z = $-3.72$, $p<0.001$). Our results confirm that TL is clearly a better choice than the classical SD scenario when we want to train one model for each subject.

%


\section{Conclusion}
\label{sec:con}

The aim of this work was to propose a deep learning architecture to model the relationship between speech stimulus and EEG. More specifically, the goal was to design a model (classifier) that can determine whether a given pair of (EEG, speech envelope) correspond to each other or not. To this end, we introduced an LSTM-based architecture which transforms small temporal segments of EEG and speech envelope to a common embedding space. In this common space, vectors of matched (EEG, envelope) should ideally be similar (aligned), while vectors of mismatched (EEG, envelope) should be dissimilar (not aligned). The reason behind using LSTM in the path of speech envelope is to give our network the ability to compensate for the brain response delay and hence synchronize with the recorded EEG. By doing this, our network finds the appropriate time delay in an automatic way from the training data, hence solving the fixed time delay problem that exists in linear models.

We compared the classification performance of the proposed LSTM-based model with that of the linear baseline and the state of the art (SoA) using data of 90 subjects. Our results show that the LSTM model significantly outperforms the other two models. We found that transferring knowledge from the generic model and fine tuning it on each subject (TL scenario) significantly increased the classification accuracy (up to 85\%) compared to the subject dependent scenario. As a result, if one intends to train one model per subject as in the SD scenario, one better uses the TL scenario instead. We conclude that the proposed LSTM architecture can be used to model the relationship between speech stimulus and EEG (also in AAD tasks) as a better alternative for the linear model. Furthermore, the learned embedded representations of EEG and speech envelope could be used in other studies as a rich representation of EEG and speech.

\bibliographystyle{IEEEbib}
\bibliography{refs}

\begin{thebibliography}{10}

\bibitem{ahissar_speech_2001}
E.~Ahissar, S.~Nagarajan, M.~Ahissar, A.~Protopapas, H.~Mahncke, and M.~M.
  Merzenich,
\newblock ``Speech comprehension is correlated with temporal response patterns
  recorded from auditory cortex,''
\newblock {\em Proceedings of the National Academy of Sciences}, vol. 98, no.
  23, pp. 13367--13372, Nov. 2001.

\bibitem{aiken_human_2008}
Steven~J. Aiken and Terence~W. Picton,
\newblock ``Human {Cortical} {Responses} to the {Speech} {Envelope},''
\newblock {\em Ear and Hearing}, vol. 29, no. 2, pp. 139, Apr. 2008.

\bibitem{ding_emergence_2012}
Nai Ding and Jonathan~Z. Simon,
\newblock ``Emergence of neural encoding of auditory objects while listening to
  competing speakers,''
\newblock {\em Proceedings of the National Academy of Sciences}, vol. 109, no.
  29, pp. 11854--11859, July 2012.

\bibitem{crosse_multivariate_2016}
Michael~J. Crosse, Giovanni~M. Di~Liberto, Adam Bednar, and Edmund~C. Lalor,
\newblock ``The {Multivariate} {Temporal} {Response} {Function} ({mTRF})
  {Toolbox}: {A} {MATLAB} {Toolbox} for {Relating} {Neural} {Signals} to
  {Continuous} {Stimuli},''
\newblock {\em Frontiers in Human Neuroscience}, vol. 10, 2016.

\bibitem{vanthornhout_speech_2018}
Jonas Vanthornhout, Lien Decruy, Jan Wouters, Jonathan~Z. Simon, and Tom
  Francart,
\newblock ``Speech {Intelligibility} {Predicted} from {Neural} {Entrainment} of
  the {Speech} {Envelope},''
\newblock {\em Journal of the Association for Research in Otolaryngology}, vol.
  19, no. 2, pp. 181--191, Apr. 2018.

\bibitem{diliberto_low-frequency_2015}
Giovanni M. Di Liberto, James A. O’Sullivan, and Edmund C. Lalor,
\newblock ``Low-{Frequency} {Cortical} {Entrainment} to {Speech} {Reflects}
  {Phoneme}-{Level} {Processing},''
\newblock {\em Current Biology}, vol. 25, no. 19, pp. 2457--2465, Oct. 2015.

\bibitem{lesenfants_predicting_2019}
D.~Lesenfants, J.~Vanthornhout, E.~Verschueren, L.~Decruy, and T.~Francart,
\newblock ``Predicting individual speech intelligibility from the cortical
  tracking of acoustic- and phonetic-level speech representations,''
\newblock {\em Hearing Research}, vol. 380, pp. 1--9, Sept. 2019.

\bibitem{mirkovic_decoding_2015}
Bojana Mirkovic, Stefan Debener, Manuela Jaeger, and Maarten~De Vos,
\newblock ``Decoding the attended speech stream with multi-channel {EEG}:
  implications for online, daily-life applications,''
\newblock {\em Journal of Neural Engineering}, vol. 12, no. 4, pp. 046007, June
  2015.

\bibitem{ding_adaptive_2013}
Nai Ding and Jonathan~Z. Simon,
\newblock ``Adaptive {Temporal} {Encoding} {Leads} to a
  {Background}-{Insensitive} {Cortical} {Representation} of {Speech},''
\newblock {\em Journal of Neuroscience}, vol. 33, no. 13, pp. 5728--5735, Mar.
  2013.

\bibitem{de_cheveigne_decoding_2018}
Alain de~Cheveigné, Daniel D.~E. Wong, Giovanni~M. Di~Liberto, Jens
  Hjortkjær, Malcolm Slaney, and Edmund Lalor,
\newblock ``Decoding the auditory brain with canonical component analysis,''
\newblock {\em NeuroImage}, vol. 172, pp. 206--216, May 2018.

\bibitem{fuglsang_noise-robust_2017}
Søren~Asp Fuglsang, Torsten Dau, and Jens Hjortkjær,
\newblock ``Noise-robust cortical tracking of attended speech in real-world
  acoustic scenes,''
\newblock {\em NeuroImage}, vol. 156, pp. 435--444, Aug. 2017.

\bibitem{de_taillez_machine_2017}
Tobias de~Taillez, Birger Kollmeier, and Bernd~T. Meyer,
\newblock ``Machine learning for decoding listeners’ attention from
  electroencephalography evoked by continuous speech,''
\newblock {\em European Journal of Neuroscience}, vol. 0, no. 0, Dec. 2017.

\bibitem{deckers_eeg-based_2018}
Lucas Deckers, Neetha Das, Amir~Hossein Ansari, Alexander Bertrand, and Tom
  Francart,
\newblock ``{EEG}-based detection of the attended speaker and the locus of
  auditory attention with convolutional neural networks,''
\newblock {\em bioRxiv}, p. 475673, Dec. 2018.

\bibitem{ciccarelli_comparison_2019}
Gregory Ciccarelli, Michael Nolan, Joseph Perricone, Paul~T. Calamia, Stephanie
  Haro, James O’Sullivan, Nima Mesgarani, Thomas~F. Quatieri, and
  Christopher~J. Smalt,
\newblock ``Comparison of {Two}-{Talker} {Attention} {Decoding} from {EEG} with
  {Nonlinear} {Neural} {Networks} and {Linear} {Methods},''
\newblock {\em Scientific Reports}, vol. 9, no. 1, pp. 11538, Dec. 2019.

\bibitem{hochreiter_long_1997}
Sepp Hochreiter and Jürgen Schmidhuber,
\newblock ``Long {Short}-{Term} {Memory},''
\newblock {\em Neural Computation}, vol. 9, no. 8, pp. 1735--1780, Nov. 1997.

\bibitem{luts2014development}
Heleen Luts, Sofie Jansen, Wouter Dreschler, and Jan Wouters,
\newblock ``Development and normative data for the flemish/dutch matrix test,''
\newblock 2014.

\bibitem{francart_apex_2008}
Tom Francart, Astrid van Wieringen, and Jan Wouters,
\newblock ``{APEX} 3: a multi-purpose test platform for auditory psychophysical
  experiments,''
\newblock {\em Journal of Neuroscience Methods}, vol. 172, no. 2, pp. 283--293,
  July 2008.

\bibitem{biesmans_auditory-inspired_2017}
W.~Biesmans, N.~Das, T.~Francart, and A.~Bertrand,
\newblock ``Auditory-{Inspired} {Speech} {Envelope} {Extraction} {Methods} for
  {Improved} {EEG}-{Based} {Auditory} {Attention} {Detection} in a {Cocktail}
  {Party} {Scenario},''
\newblock {\em IEEE Transactions on Neural Systems and Rehabilitation
  Engineering}, vol. 25, no. 5, pp. 402--412, May 2017.

\bibitem{somers_generic_2018}
Ben Somers, Tom Francart, and Alexander Bertrand,
\newblock ``A generic {EEG} artifact removal algorithm based on the
  multi-channel {Wiener} filter,''
\newblock {\em Journal of Neural Engineering}, vol. 15, no. 3, pp. 036007, Feb.
  2018.

\bibitem{osullivan_attentional_2015}
James~A. O'Sullivan, Alan~J. Power, Nima Mesgarani, Siddharth Rajaram, John~J.
  Foxe, Barbara~G. Shinn-Cunningham, Malcolm Slaney, Shihab~A. Shamma, and
  Edmund~C. Lalor,
\newblock ``Attentional {Selection} in a {Cocktail} {Party} {Environment} {Can}
  {Be} {Decoded} from {Single}-{Trial} {EEG},''
\newblock {\em Cerebral Cortex (New York, N.Y.: 1991)}, vol. 25, no. 7, pp.
  1697--1706, July 2015.

\end{thebibliography}

\end{document}